\documentstyle[prd,aps,preprint,psfig]{revtex}

\begin{document}
\newcommand{\keV}{~\mbox{keV}}
\newcommand{\MeV}{~\mbox{MeV}}
\newcommand{\GeV}{~\mbox{GeV}}
\tighten
\preprint{\begin{tabular}{l}
\hbox to\hsize{\hfill UT-845}\\
\hbox to\hsize{\hfill RESCEU-6/99}\end{tabular}}
\bigskip
\title{Superheavy Dark Matter and Thermal Inflation}
\author{T. Asaka$^1$, M. Kawasaki$^2$, and T. Yanagida$^{1,2}$}
\address{$^1$Department of Physics,  University of
  Tokyo, Tokyo 113-0033, Japan}
\address{$^2$Research Center for the Early Universe,
  University of Tokyo, Tokyo 113-0033, Japan}
\date{April 23, 1999}

\maketitle
\begin{abstract}
The thermal inflation is the most plausible mechanism
that solves the cosmological moduli problem naturally.
We discuss relic abundance of superheavy particle $X$
in the presence of the thermal inflation
assuming that its lifetime is longer than the age of the universe,
and show that the long-lived particle $X$ of mass 
$10^{12}$--$10^{14}$ GeV may form a part of the dark matter 
in the present universe
in a wide region of parameter space of the thermal inflation model.
The superheavy dark matter of mass $\sim 10^{13}$ GeV
may be interesting in particular,
since its decay may account for the observed ultra high-energy
cosmic rays if the lifetime of the $X$ particle is sufficiently long.
\end{abstract}

\clearpage
%
A large class of string theories predicts a number of flat directions,
called moduli fields $\phi$ \cite{Green-Schwartz-Witten}.
They are expected to acquire their masses of order of the 
gravitino mass $m_{3/2}$ from some nonperturbative effects
of supersymmetry (SUSY) breaking \cite{Carlos-Casas-Quevedo-Roulet}.
The gravitino mass lies in a range of $10^{-2}$ keV--1 GeV
for gauge-mediated SUSY breaking models \cite{GMSB}
and in a range of 100 GeV--1 TeV for hidden-sector SUSY breaking
models \cite{HSSB}.
It is well known \cite{Moduli-Problem,Carlos-Casas-Quevedo-Roulet} 
that such moduli fields are produced too much as 
coherent oscillations in the early universe
and conflict with various cosmological observations.
Therefore, we must invoke a mechanism such as late-time 
entropy production to dilute the moduli density substantially.

The thermal inflation \cite{Lyth-Stewart} is the most plausible
mechanism to produce an enormous amount of entropy at the late time
of the universe's evolution.
In recent articles \cite{Asaka-Hashiba-Kawasaki-Yanagida} 
we have shown that
the thermal inflation is very successful in solving the above 
cosmological moduli problem.
Since it produces a tremendous amount of entropy to dilute the moduli 
density, abundances of any relic particles
are substantially diluted simultaneously,
which may provide a new possibility for a superheavy $X$ particle
to be a part of the dark matter in the universe.%
\footnote{
Other candidates for the dark matter 
in the presence of the thermal inflation are
the moduli themselves whose masses are less than about 100 keV 
\cite{Asaka-Hashiba-Kawasaki-Yanagida,%
Asaka-Hashiba-Kawasaki-Yanagida2},
and the axion with a relatively high decay constant 
$f_a \sim 10^{15}$--$10^{16}$ GeV
\cite{Kawasaki-Yanagida-axion,axionDM}.
}
In this paper we show that it is indeed the case
if the mass of $X$ particle is of order of $10^{12}$--$10^{14}$ GeV
and its lifetime is longer than the age of the universe.
Such a long-lived $X$ particle is particularly interesting,
since its decay may naturally explain 
\cite{Berezinsky-Kachelriess-Vilenkin,Kuzmin-Ruvakov,%
Kuzmin-Tkachev,Benakli-Ellis-Nanopoulos,Birkel-Sarkar} 
the observed ultra high-energy cosmic
rays \cite{UHCR1,UHCR2,UHCR3} beyond the Greisen-Zatsepin-Kuzmin cutoff
\cite{GZKcutoff} when the lifetime is sufficiently long.

In this paper we consider that the particle $X$ was primordially 
in the thermal equilibrium%
\footnote{
For the superheavy particle $X$ to be in the thermal equilibrium,
the reheating temperature after the primordial inflation
should be higher than $m_X \sim 10^{13}$ GeV.
Such a high reheating temperature is realized in e.g.
hybrid inflation models \cite{Hybrid-Inf}.
Although a large number of gravitinos are also produced in
this case,
they are sufficiently diluted by the thermal inflation
and become harmless.
}
and froze out at the cosmic temperature $T = T_f$ 
when it was nonrelativistic $x_f = m_X/T_f > 1$
(i.e., the $X$ is left as a cold relic.).%
\footnote{
Nonthermal productions of superheavy particles
were discussed in Refs.
\cite{Kuzmin-Tkachev,%
Chung-Kolb-Riotto}.
}
Then, the present relic abundance of $X$ 
(the ratio of the present energy density of $X$
to the critical density $\rho_{cr}$) is estimated by using the 
thermally-averaged annihilation cross section of the $X$,
$\langle \sigma_X \left| v \right| \rangle$, as\cite{Kolb-Turner}
\begin{eqnarray}
    \Omega_X^0 h^2 
    = 
    \frac{ 0.76 (n_f+1) x_f }
    { g_\ast (T_f)^{\frac 12}  M_G ( h^{-2} \rho_{cr} / s_0 )
      \langle \sigma_X \left| v \right| \rangle }
\end{eqnarray}
where $g_{\ast} (T_f) \simeq 200$ 
counts the effective degrees of freedom at $T=T_f$,
$M_G \simeq 2.4 \times 10^{18}$ GeV denotes the reduced Planck scale,
and $s_0$ is the present entropy density 
$(\rho_{cr}/s_0) \simeq 3.6 \times 10^{-9} h^2$ GeV
with the present Hubble parameter $h$ in units of 
100 km/sec$\cdot$Mpc$^{-1}$. 
Here $n_f$ parametrizes the dependence on $T$ of 
the annihilation cross section
and $n_f=0$ for $s$-wave annihilation, 
$n_f=1$ for $p$-wave annihilation, etc.
Assuming the $s$-wave annihilation with
$\langle \sigma_X \left| v \right| \rangle \simeq m_X^{-2}$
($x_f \simeq 7$ for $m_X \sim 10^{13}$ GeV) we obtain
\begin{eqnarray}
    \label{OmX}
    \Omega_X^0 h^2 \simeq
    \frac{ 0.4 m_X^2 }{ M_G ( h^{-2} \rho_{cr} / s_0 ) }
    =
    4.3 \times 10^{15} \left( \frac{ m_X }{ 10^{13} \GeV } \right)^{2}.
\end{eqnarray}
Therefore, the superheavy particle $X$ of $m_X \gtrsim 10^{5}$ GeV 
leads to overclosure 
of the universe if its lifetime is longer than the age of 
the universe.
In order to realize $\Omega_X h^2 \lesssim 1$
the dilution factor more than about $10^{16}$
is required for $m_X \sim 10^{13}$ GeV for example.
However, such a huge dilution 
may be naturally provided by the thermal inflation. 
In this paper, 
we examine whether 
the thermal inflation could sufficiently
reduce the energy density $\Omega_X$ 
of the superheavy particle even if it was in the thermal 
equilibrium, as well as that of the string moduli.

Let us start with reviewing briefly the thermal inflation model
\cite{Lyth-Stewart,Asaka-Hashiba-Kawasaki-Yanagida}.
The potential of the inflaton field $S$ is given by
\begin{eqnarray}
    \label{V_S}
    V = V_0 - m_0^2 \left| S \right|^2
    + \frac{ \left| S \right|^{2n+4} }{ M_\ast^{2n} },
\end{eqnarray}
where $-m_0^2 $ denotes a soft SUSY breaking negative mass 
squared which is expected to be of order of the electroweak scale.
$M_\ast$ denotes the cutoff scale 
and $V_0$ is the vacuum energy.
Then the vacuum expectation value of $S$ is estimated as 
\begin{eqnarray}
    \label{VEV-S}
    \langle S \rangle \equiv M
    =
    \left( n+2 \right)^{ - \frac{1}{2(n+1)} }
    \left( m_0 M_\ast^n \right)^{ \frac{ 1 }{ n+1 } },
\end{eqnarray}
and $V_0$ is fixed as
\begin{eqnarray}
    V_0 = \frac{ n+1 }{ n+2 } m_0^2 M^2,
\end{eqnarray}
so that the cosmological constant vanishes at the true vacuum.
The inflaton $\sigma$ ($\equiv$Re$S$), which we call it a flaton,
obtains a mass $m_\sigma^2 = 2 (n+1) m_0^2$.

After the thermal inflation ends, 
the vacuum energy is transferred into the thermal bath
through flaton decay and increases the entropy of the universe
by a factor \cite{Lyth-Stewart,Asaka-Hashiba-Kawasaki-Yanagida}%
\footnote{
Here we have assumed that 
the flaton can not decay into two $R$-axions which is 
an imaginary part of $S$ to obtain a successful dilution.
See Ref. \cite{Asaka-Hashiba-Kawasaki-Yanagida}.
}
\begin{eqnarray}
    \Delta 
    \simeq 
    \frac{ V_0 }{ 66 m_\sigma^3 T_R }.
\end{eqnarray}
Here $T_R$ is the reheating temperature after the thermal inflation
and determined by the flaton decay width $\Gamma_\sigma$
which can be written as
\begin{eqnarray}
    \Gamma_\sigma = C_\sigma \frac{ m_\sigma^3 }{ M^2 },
\end{eqnarray}
where $C_\sigma$ is a dimension-less parameter depending on 
the decay modes.
We consider that the flaton $\sigma$ dominantly decays into 
two photons or two gluons, and 
$C_\sigma$ is given by%
\footnote{
We assume that the decay amplitudes are obtained from one-loop
diagrams of some heavy particles. 
See also Ref. \cite{Asaka-Hashiba-Kawasaki-Yanagida}.
}
\begin{eqnarray}
    C_\sigma \simeq
    \left\{
        \begin{array}[]{ll}
        \frac{1}{4\pi} 
        \left( \frac{ \alpha }{ 4 \pi } \right)^2 
        &
        \mbox{for} ~~\sigma \rightarrow \gamma \gamma
        \\
        \frac{1}{4\pi} 
        \left( \frac{ \alpha_s }{ 4 \pi } \right)^2 
        &
        \mbox{for} ~~\sigma \rightarrow g g
    \end{array}
    \right..
\end{eqnarray}
Then the reheating temperature is estimated as
\begin{eqnarray}
    \label{TR}
    T_R 
    =
    \left( \frac{ 90 }{ \pi^2 g_\ast (T_R) } \right)^{1 \over 4 }
    \sqrt{ \Gamma_\sigma M_G }
    \simeq
    0.96 ~ C_\sigma^{1 \over 2 }
    \frac{ m_\sigma^{3 \over 2} M_G^{ 1 \over 2} }{ M }.
\end{eqnarray}
The entropy-production factor $\Delta$, therefore, can be written as
\begin{eqnarray}
    \Delta 
    &=&
    \frac{ M^3 }
         { 130 (n+2) C_\sigma^{ 1 \over 2} 
           m_\sigma^{ 5 \over 2 } M_G^{ 1 \over 2 } },
\end{eqnarray}
and for $n=1$ we obtain 
\begin{eqnarray}
    \label{Delta}
    \Delta 
    &\simeq&
    \left\{
        \begin{array}[]{ll}
            1.0 \times 10^{17} ~
            \displaystyle{
            \left( \frac{ m_\sigma }{100 \GeV} \right)^{ - \frac 52}
            \left( \frac{ M }{ 10^{10} \GeV } \right)^3
            }
            &
            \mbox{for} ~~\sigma \rightarrow \gamma \gamma
            \\
            6.4 \times 10^{15} ~
            \displaystyle{
            \left( \frac{ m_\sigma }{100 \GeV} \right)^{ - \frac 52}
            \left( \frac{ M }{ 10^{10} \GeV } \right)^3
            }
            &
            \mbox{for} ~~\sigma \rightarrow g g
        \end{array}
    \right..
\end{eqnarray}
Here the values $m_\sigma = 100$ GeV and $M = 10^{10}$ GeV 
correspond to  $M_\ast = 3.5 \times 10^{18}$ GeV for $n=1$.
In the following analysis, we take $n=1$ for simplicity.
From Eq. (\ref{Delta}) we see that the thermal inflation 
can dilute relic particle density extensively by producing 
an enormous entropy.
Here it should be noted that the reheating temperature 
should be $T_R \gtrsim 1$ MeV to keep 
the big bang nucleosynthesis successful \cite{Kawasaki-Kohri-Sugiyama}%
\footnote{
In Ref. \cite{Kawasaki-Kohri-Sugiyama}
the lower bound on the reheating temperature is 
determined as about 0.5 MeV.
Since our definition of the reheating temperature is different
by a factor $\sqrt{3}$, 
it leads to $T_R \gtrsim 1$ MeV in our case.
},
which leads to the upper bounds on $M$ from Eq. (\ref{TR}) as
\begin{eqnarray}
    M 
    &=&
    0.96 ~ C_\sigma^{\frac 12} 
    \frac{ m_\sigma^{\frac 32} M_G^{\frac 12} }{ T_R }
    \lesssim
    \left\{
        \begin{array}[]{ll}
            2.4 \times 10^{11} \GeV~
            \displaystyle{
            \left( \frac{ m_\sigma }{100 \GeV} \right)^{ \frac 32}
            }
            &
            \mbox{for} ~~\sigma \rightarrow \gamma \gamma
            \\
            3.9 \times 10^{12} \GeV~
            \displaystyle{
            \left( \frac{ m_\sigma }{100 \GeV} \right)^{ \frac 32}
            }
            &
            \mbox{for} ~~\sigma \rightarrow g g
        \end{array}
    \right..
\end{eqnarray}
These are translated into the upper bounds on $M_\ast$ as
\begin{eqnarray}
    \label{UB-Mst}
    M_\ast
    &=&
    3.2 ~ C_\sigma
    \frac{ m_\sigma^2 M_G }{ T_R^2 }
    \lesssim
    \left\{
        \begin{array}[]{ll}
            2.1 \times 10^{21} \GeV~
            \displaystyle{
            \left( \frac{ m_\sigma }{100 \GeV} \right)^{ 2 }
            }
            &
            \mbox{for} ~~\sigma \rightarrow \gamma \gamma
            \\
            5.4 \times 10^{23} \GeV~
            \displaystyle{
            \left( \frac{ m_\sigma }{100 \GeV} \right)^{ 2 }
            }
            &
            \mbox{for} ~~\sigma \rightarrow g g
        \end{array}
    \right..
\end{eqnarray}

\begin{figure}[t]
    \centerline{\psfig{figure=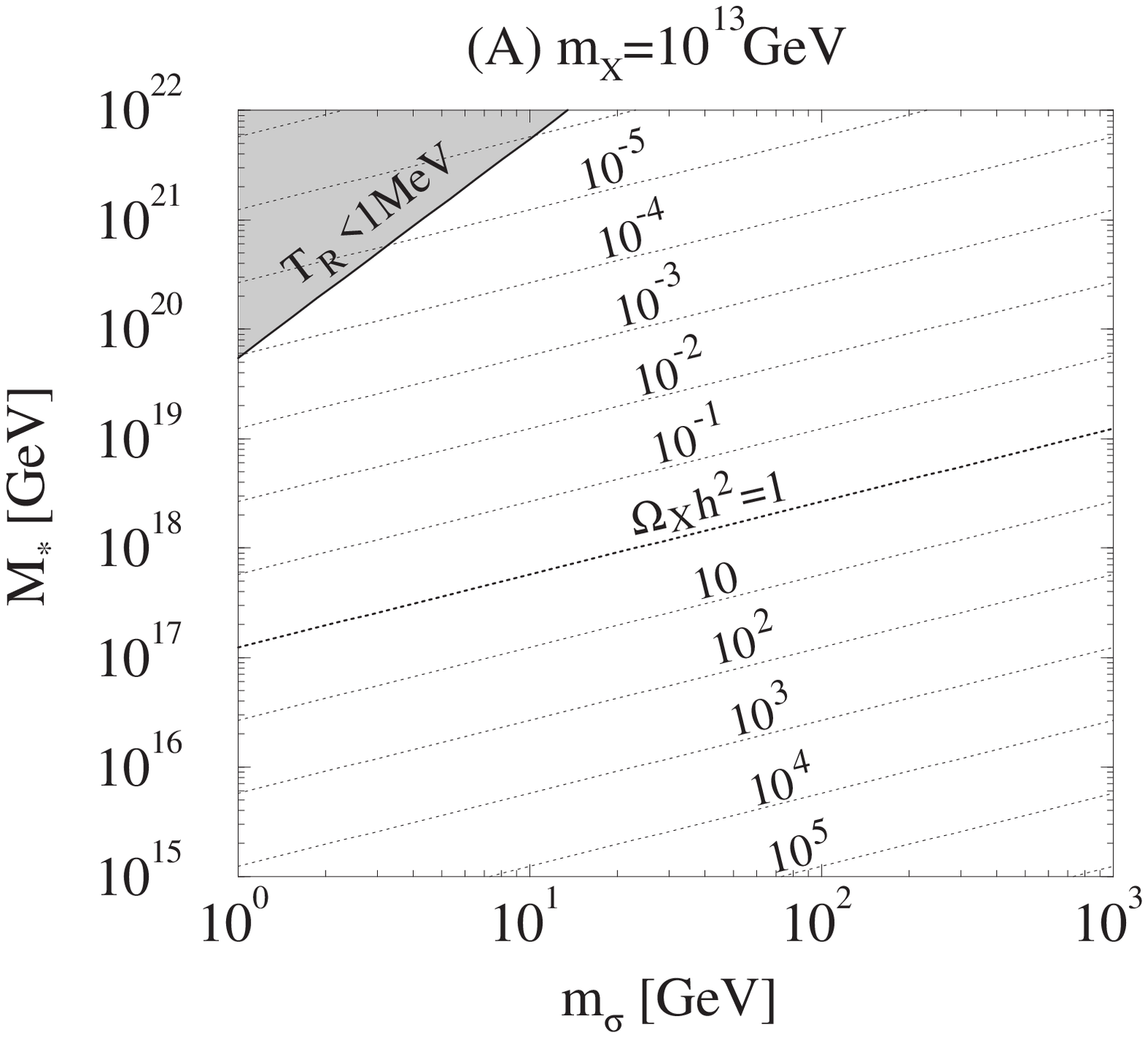,width=10cm}}
    \centerline{\psfig{figure=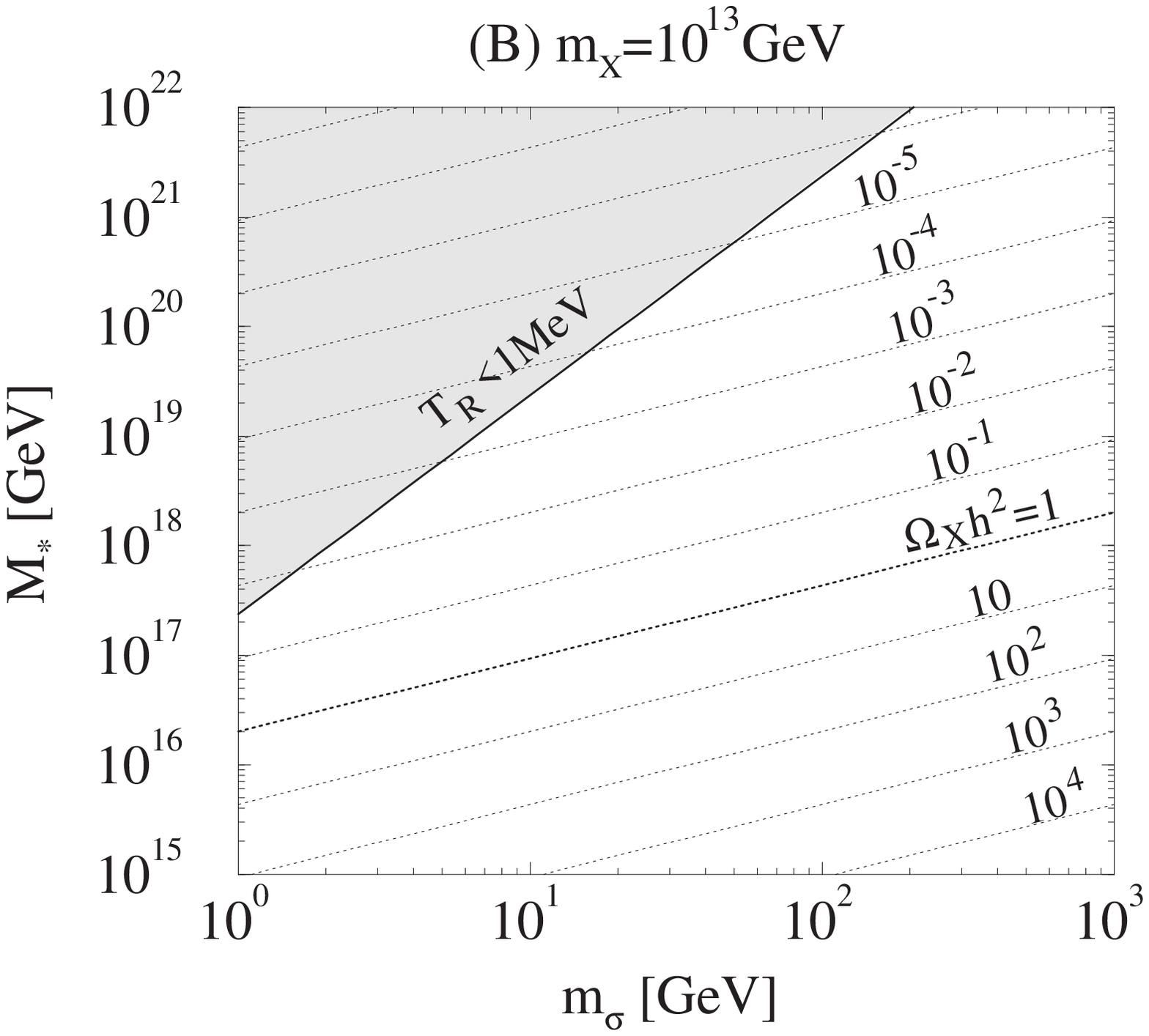,width=10cm}}
\caption{
The contour lines of the relic abundance ($\Omega_X h^2$)
of the superheavy particle $X$ of mass $m_X = 10^{13}$ GeV
in the presence of the thermal inflation with $n=1$ 
for the cases that a flaton $\sigma$ decays mainly into two gluons (A)
and that a flaton $\sigma$ decays mainly into two photons (B).
The contour lines are represented by the dotted lines.
(The contour line of $\Omega_X h^2$=1 is represented by the thick
dotted line.)
We show the upper bound on the cutoff scale $M_\ast$ obtained
from $T_R >$ 1 MeV by the thick solid line.
}
    \label{fig:OmX}
\end{figure}

In the presence of the thermal inflation 
the relic abundance of the superheavy particle $X$ 
[Eq. (\ref{OmX})] is reduced by the factor $\Delta$ 
in Eq. (\ref{Delta}) as
\begin{eqnarray}
    \label{OmX-TI}
    \Omega_X h^2 
    &=&
    \Omega_X^0 h^2 \times \frac{ 1 }{ \Delta }
    =
    140 ~C_\sigma^{ \frac 12}
     \frac{ m_X^2 m_\sigma^{\frac 52} }
          { M^3 M_G^{\frac 12} ( h^{-2} \rho_{cr}/s_0) }
    \nonumber \\
    &\simeq&
    \left\{
        \begin{array}[]{ll}
            0.04 ~
            \displaystyle{
            \left( \frac{ m_X }{ 10^{13} \GeV } \right)^2
            \left( \frac{ m_\sigma }{100 \GeV} \right)^{ \frac 52 }
            \left( \frac{ M }{10^{10} \GeV} \right)^{ -3 }
            }
            &
            \mbox{for} ~~\sigma \rightarrow \gamma \gamma
            \\
            0.68 ~
            \displaystyle{
            \left( \frac{ m_X }{ 10^{13} \GeV } \right)^2
            \left( \frac{ m_\sigma }{100 \GeV} \right)^{ \frac 52 }
            \left( \frac{ M }{10^{10} \GeV} \right)^{ -3 }
            }
            &
            \mbox{for} ~~\sigma \rightarrow g g
        \end{array}
    \right.,
\end{eqnarray}
In Fig. \ref{fig:OmX} we show the contour of $\Omega_X$
in the parameter space of the thermal inflation model
(in $m_\sigma$-$M_\ast$ plane).
We find that 
the thermal inflation can 
naturally realize $\Omega_X h^2 \lesssim 1$ 
keeping the constraint $T_R \gtrsim 1$ MeV
in a large region of the parameter space.

As mentioned in the introduction, 
the thermal inflation was originally proposed 
as a solution to the cosmological moduli problem.
Therefore, we turn to examine whether
the thermal inflation could sufficiently dilute
not only the density of the superheavy particle $X$ 
but also that of the string moduli, simultaneously.

When the Hubble parameter becomes comparable to the moduli masses,
the moduli $\phi$ start coherent oscillations and 
the corresponding cosmic temperature is estimated as
\begin{eqnarray}
    T_\phi \simeq 7.2 \times 10^{6} \GeV
    \left( \frac{ m_\phi }{ 100 \keV } \right)^{\frac 12}.
\end{eqnarray}
Here notice that the moduli oscillations always begin 
after the $X$ freezes out since $T_f \sim m_X \gg T_\phi$ 
even for the heavy moduli $m_\phi \simeq m_{3/2} \simeq
100$ GeV--1 TeV predicted in hidden sector SUSY breaking models.
Because the initial amplitudes of the oscillations, $\phi_0$,
are expected to be $\phi_0 \sim M_G$,
the present abundances%
\footnote{
If the moduli masses are less than about 100 MeV,
the moduli become stable until the present.
On the other hand,
for the heavier moduli mass region,
$\Omega_\phi$ is
regarded as the ratio, $(\rho_\phi/s)_D/ (\rho_{cr}/s_0)$,
where $(\rho_\phi/s)_D$ denotes the ratio of 
the energy density of the moduli oscillations 
to the entropy density when the moduli decay.
}
of the moduli oscillations are given by
\begin{eqnarray}
    \label{OmBB0}
    \Omega_\phi h^2 = 2.5 \times 10^{14}
    \left( \frac{ m_\phi }{ 100 \keV} \right)^{\frac 12}
    \left( \frac{ \phi_0 }{ M_G } \right)^2.
\end{eqnarray}
Such a huge energy density of the moduli leads to cosmological
difficulties for the typical moduli mass regions 
predicted in both gauge-mediated SUSY breaking and hidden-sector
SUSY breaking scenarios.

However, if the universe experienced the thermal inflation,
the moduli abundances are reduced by the factor $\Delta$ 
[Eq. (\ref{Delta})] as \cite{Lyth-Stewart,Asaka-Hashiba-Kawasaki-Yanagida}
\begin{eqnarray}
    \label{OmBB}
    \left( \Omega_\phi \right)_{BB} h^2 
    &=&
    22 C_\sigma^{ \frac 12 }
    \frac{ m_\phi^{ \frac 12 } m_\sigma^{\frac 52} M_G }
         { M^3 ( h^{-2} \rho_{cr} / s_0 ) }
    \left( \frac{ \phi_0 }{ M_G } \right)^2
    \nonumber \\
    &\simeq&
    \left\{
        \begin{array}[]{ll}
            2.4 \times 10^{-3}~
            \displaystyle{
            \left( \frac{ m_\phi }{ 100 \keV } \right)^{ \frac 12 }
            \left( \frac{ m_\sigma }{100 \GeV} \right)^{ \frac 52 }
            \left( \frac{ M }{10^{10} \GeV} \right)^{ -3 }
            \left( \frac{ \phi_0 }{ M_G } \right)^2
            }
            &
            \mbox{for} ~~\sigma \rightarrow \gamma \gamma
            \\
            3.9 \times 10^{-2}~
            \displaystyle{
            \left( \frac{ m_\phi }{ 100 \keV } \right)^{ \frac 12 }
            \left( \frac{ m_\sigma }{100 \GeV} \right)^{ \frac 52 }
            \left( \frac{ M }{10^{10} \GeV} \right)^{ -3 }
            \left( \frac{ \phi_0 }{ M_G } \right)^2
            }
            &
            \mbox{for} ~~\sigma \rightarrow g g
        \end{array}
    \right..
\end{eqnarray}
We call these moduli produced at $T = T_\phi$ as ``big-bang'' moduli.
In deriving Eqs. (\ref{OmBB0}) and (\ref{OmBB}) we have assumed
that the energy of the universe is radiation-dominated
when the big-bang modulus starts to oscillate at $H \simeq m_{\phi}$.
This assumption is justified for
$\left( m_\phi / 100 \keV \right)^{1/2}
\left( m_X / 10^{13} \GeV \right)^{-2} \gtrsim 2.9$.
On the other hand, when the energy at $H \simeq m_\phi$ is dominated by 
the superheavy particle $X$,
the present abundance of the big-bang modulus is related to 
the abundance of $X$ (\ref{OmX-TI}) as
\begin{eqnarray}
    \left( \Omega_\phi \right)_{BB} h^2
    = 
    \frac{ 1 }{ 6} \Omega_X h^2 
    \left( \frac{ \phi_0 }{ M_G } \right)^2.
\end{eqnarray}
Therefore, both abundances are comparable for $\phi_0 \sim M_G$.

Furthermore, it should be noticed that 
the secondary oscillations of the moduli 
start just after the thermal inflation ends \cite{Lyth-Stewart}.
We call the moduli produced by these secondary oscillations
as ``thermal inflation'' moduli.
The present abundances of the thermal inflation moduli 
are estimated as \cite{Lyth-Stewart,Asaka-Hashiba-Kawasaki-Yanagida}
\begin{eqnarray}
    \label{OmTI}
    \left( \Omega_\phi \right)_{TI} h^2
    &=&
    6.0 \times 10^{-2} ~ C_\sigma^{ \frac 12}
    \frac{ m_\sigma^{ \frac 72 } M }
         { m_\phi^2 M_G^{ \frac 32} ( h^{-2} \rho_{cr} / s_0 ) }
    \left( \frac{ \phi_0 }{ M_G } \right)^2
    \nonumber \\
    &\simeq&
    \left\{
        \begin{array}[]{ll}
            7.4 ~
            \displaystyle{
            \left( \frac{ m_\phi }{ 100 \keV } \right)^{ - 2 }
            \left( \frac{ m_\sigma }{100 \GeV} \right)^{ \frac 72 }
            \left( \frac{ M }{10^{10} \GeV} \right)
            \left( \frac{ \phi_0 }{ M_G } \right)^2
            }
            &
            \mbox{for} ~~\sigma \rightarrow \gamma \gamma
            \\
            1.2 \times 10^{2}~
            \displaystyle{
            \left( \frac{ m_\phi }{ 100 \keV } \right)^{ - 2 }
            \left( \frac{ m_\sigma }{100 \GeV} \right)^{ \frac 72 }
            \left( \frac{ M }{10^{10} \GeV} \right)
            \left( \frac{ \phi_0 }{ M_G } \right)^2
            }
            &
            \mbox{for} ~~\sigma \rightarrow g g
        \end{array}
    \right..
\end{eqnarray}

We see from Eqs. (\ref{OmBB}) and (\ref{OmTI})
that the thermal inflation dilutes the moduli density substantially.
In fact, it has been shown in Refs. 
\cite{Lyth-Stewart,Asaka-Hashiba-Kawasaki-Yanagida,Asaka-Kawasaki}
that two moduli mass regions: (i) $m_\phi \lesssim 1$ MeV and
(ii) $m_\phi \gtrsim 10$ GeV
survive various cosmological constraints 
in the presence of the thermal inflation.

\begin{figure}[t]
    \centerline{\psfig{figure=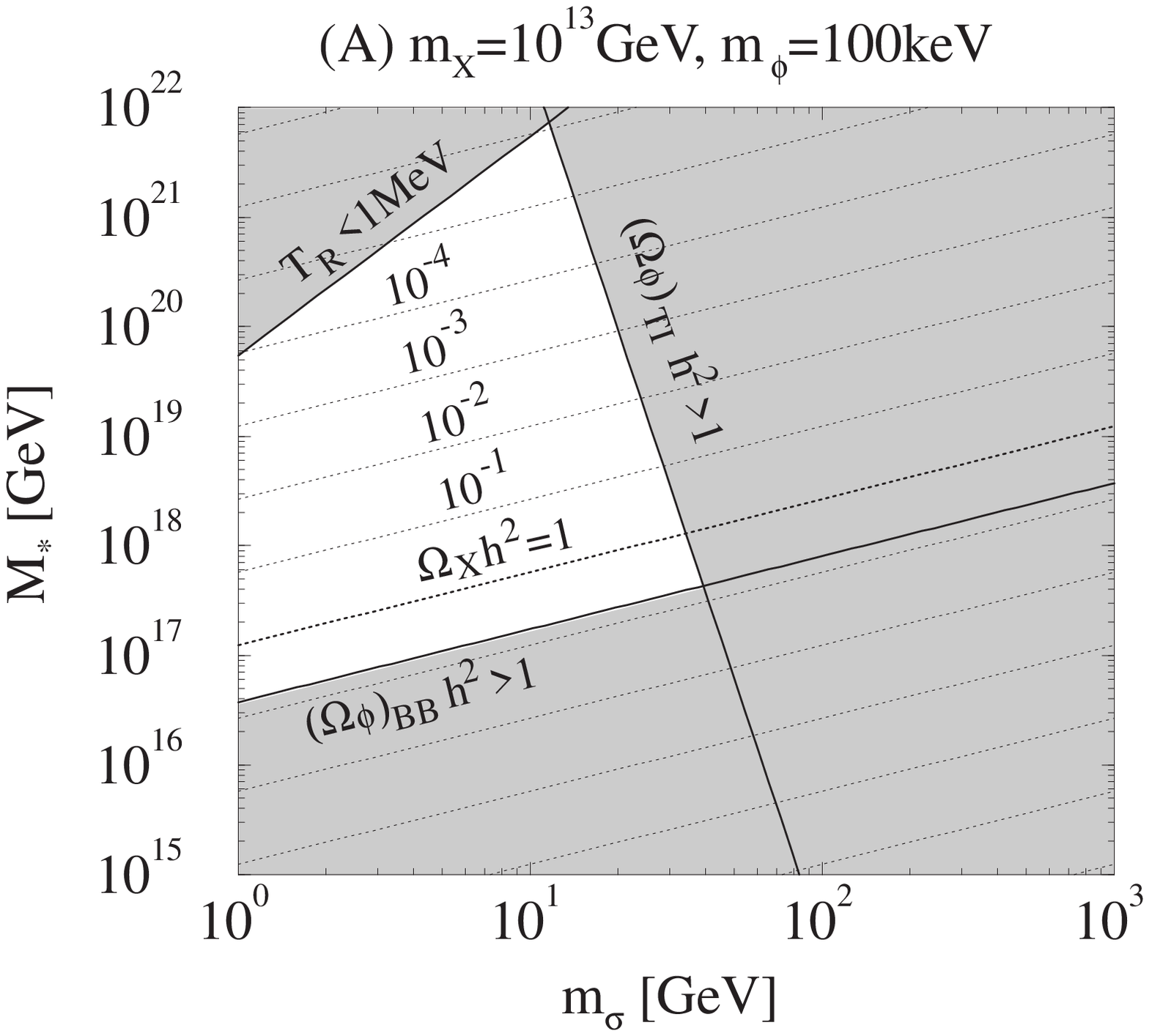,width=10cm}}
    \centerline{\psfig{figure=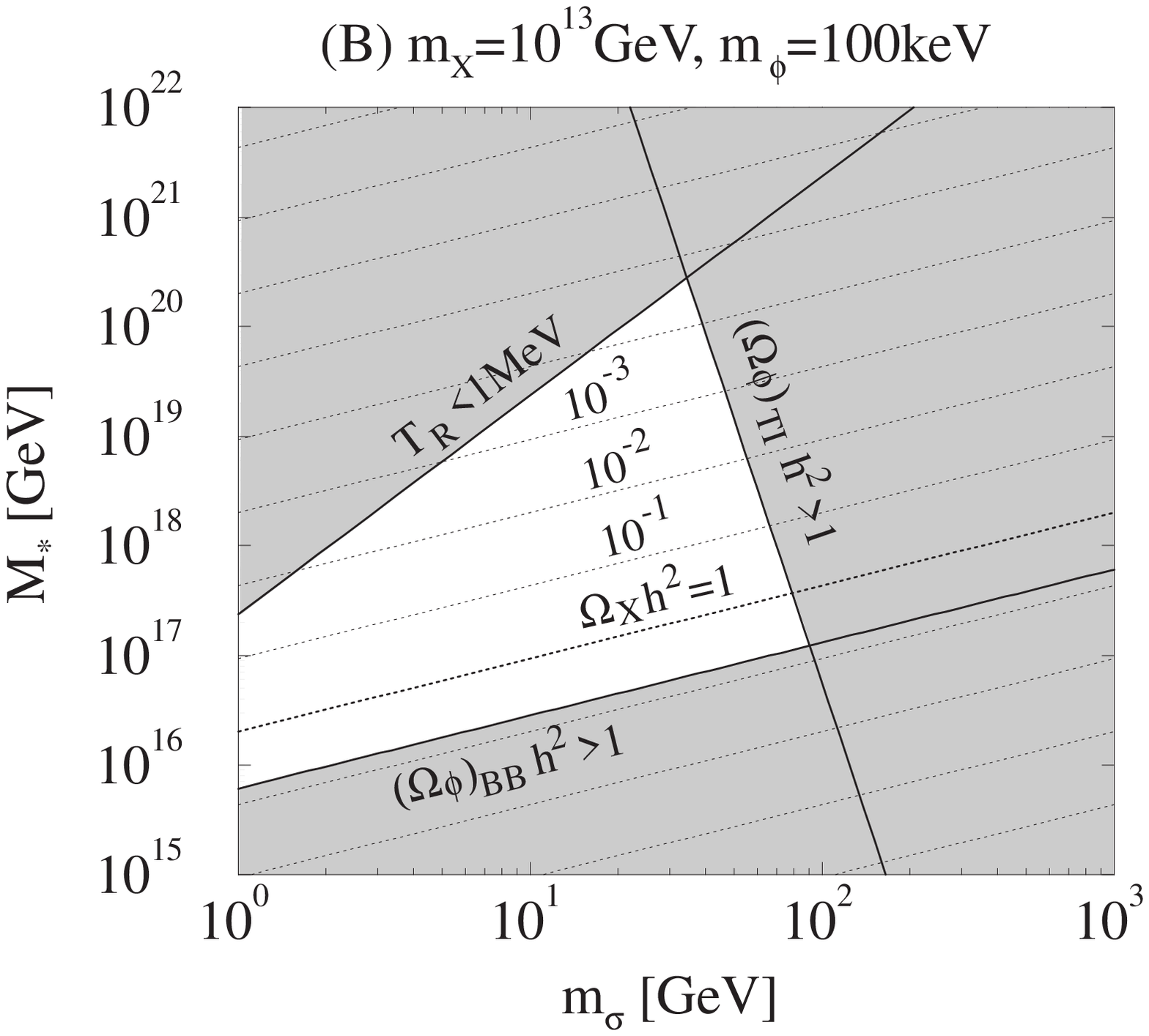,width=10cm}}
\caption{
The contour lines of the relic abundance ($\Omega_X h^2$)
of the superheavy particle $X$ of mass $m_X = 10^{13}$ GeV
in the presence of the thermal inflation 
with $n=1$ for the cases that 
a flaton $\sigma$ decays mainly into two gluons (A)
and that a flaton $\sigma$ decays mainly into two photons (B).
The contour lines are represented by the dotted lines.
(The contour line of $\Omega_X h^2$=1 is represented by the thick
dotted line.)
We show the upper bound on the cutoff scale $M_\ast$ obtained
from $T_R >$ 1 MeV by the thick solid line.
The lower bound on $M_\ast$ from $(\Omega_\phi)_{BB} h^2 < 1$
and the upper bound on $M_\ast$ from $(\Omega_\phi)_{TI} h^2 < 1$
for the case $m_\phi = 100$ keV
are also shown by the thick solid lines.
}
    \label{fig:OmXmp100keV}
\end{figure}

We are now at the point to examine whether the thermal inflation can
solve the moduli problem and 
also dilute sufficiently the superheavy particle $X$ at the same time.
First, we consider the lighter allowed region of the moduli masses
predicted in gauge-mediated SUSY breaking models.
In Fig. \ref{fig:OmXmp100keV} 
we show the contour plot of the abundance of $X$ 
of mass $m_X = 10^{13}$ GeV as well as the various constraints
in the $m_\sigma$-$M_\ast$ plane for $m_\phi = 100$ keV.
Such light moduli are constrained from the overclosure limit.%
\footnote{
Notice that for the modulus of mass $m_\phi = 100$ keV 
the constraint from the diffuse x$(\gamma)$-ray backgrounds
is not so severe as the overclosure limit.
It gives a more stringent upper bound on the modulus abundance
for 100 MeV $\gtrsim m_\phi \gtrsim 200$ keV
\cite{Kawasaki-Yanagida-xray}.
}
Then the requirements $(\Omega_\phi)_{BB} h^2 \lesssim$ 1
and $(\Omega_\phi)_{TI} h^2 \lesssim$ 1 put 
the lower and upper bounds on $M_\ast$ respectively.
Furthermore, the condition $T_R \gtrsim 1$ MeV leads to 
the upper bound on $M_\ast$ as represented in Eq. (\ref{UB-Mst}).
We see from the figures that
in the parameter space which survives above constraints
one can obtain $\Omega_X h^2 \sim 10^{-6}$--$1$ for $m_X = 10^{13}$ GeV:
i.e.,
the thermal inflation can dilute the abundance of the superheavy 
particle $X$ sufficiently.
Similar results are obtained in the allowed moduli mass 
region $m_\phi \simeq 10^{-2}$ keV--$1$ MeV.
For example, when $m_\phi = 10^{-2}$ keV,
the thermal inflation solves the moduli problem
in the parameter regions $m_\sigma \simeq 10^{-2}$ GeV
--$1$ GeV and $M_\ast \simeq 10^{13}$ GeV--$10^{17}$ GeV
and in this region we obtain
$\Omega_X h^2 \sim 10^{-1}$--$1$ for $m_X = 10^{13}$ GeV.
We find that the thermal inflation is very successful,
in a whole range of $m_\phi \simeq 10^{-2}$ keV--1 MeV,
to solve the moduli problem as well as 
to reduce the density of 
the superheavy particle whose mass is $m_X \sim 10^{13}$ GeV
even if it was in the thermal bath in the early universe.

\begin{figure}[t]
    \centerline{\psfig{figure=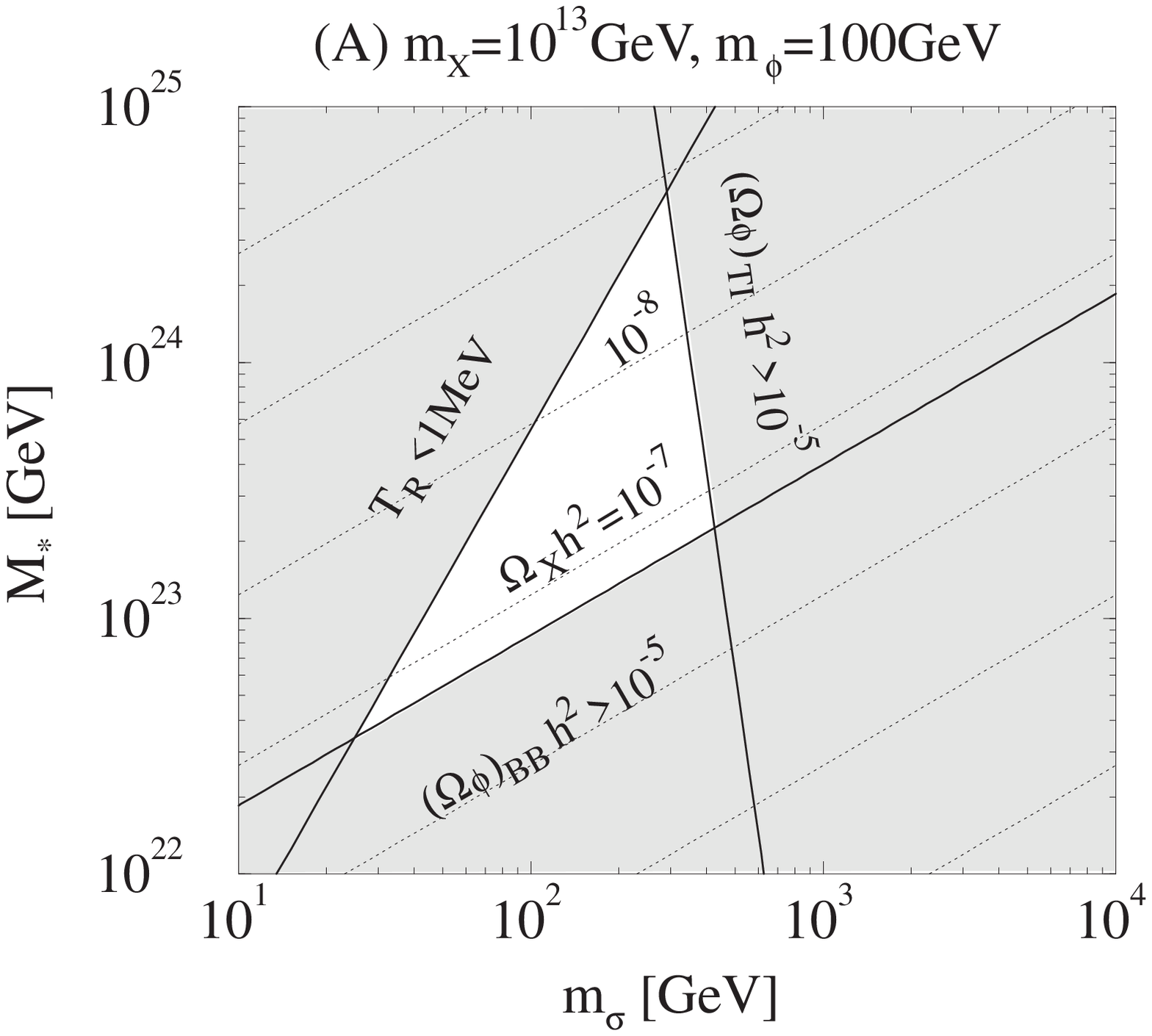,width=10cm}}
    \centerline{\psfig{figure=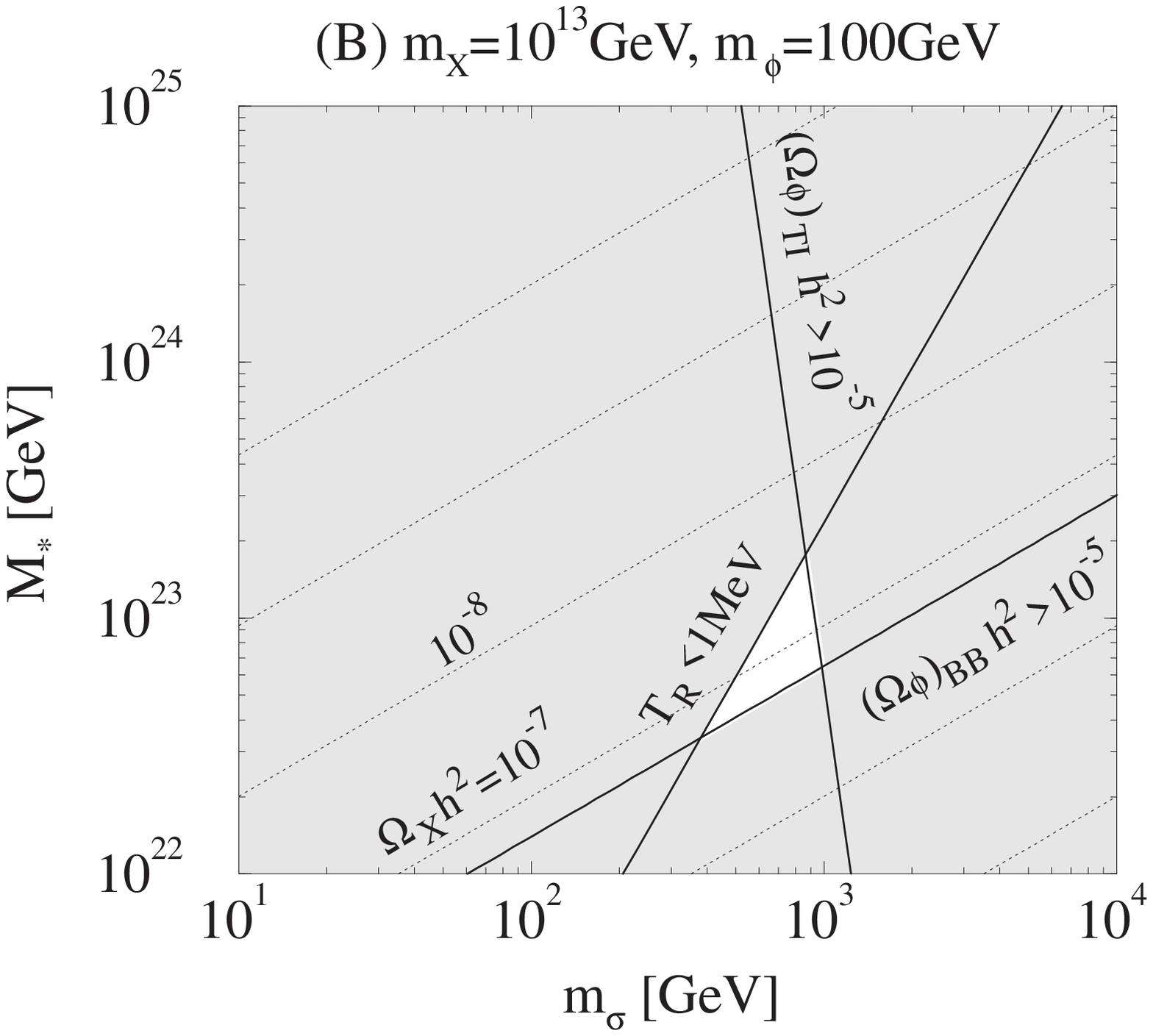,width=10cm}}
\caption{
The contour lines of the relic abundance ($\Omega_X h^2$)
of the superheavy particle $X$ of mass $m_X = 10^{13}$ GeV
in the presence of the thermal inflation
with $n=1$ for the cases that a flaton $\sigma$ 
decays mainly into two gluons (A) and that
a flaton $\sigma$ decays mainly into two photons (B).
The contour lines are represented by the dotted lines.
We show the upper bound on the cutoff scale $M_\ast$ 
obtained from $T_R >$ 1 MeV by the thick solid line.
The lower bound on $M_\ast$ from $(\Omega_\phi)_{BB} h^2< 10^{-5}$
and the upper bound on $M_\ast$ from $(\Omega_\phi)_{TI} h^2< 10^{-5}$
for the case $m_\phi = 100$ GeV
are also shown by the thick solid lines.
}
    \label{fig:OmXmp100GeV}
\end{figure}

Next, we turn to discuss the case of the heavier moduli of masses
$m_\phi \simeq 100$ GeV--$1$ TeV predicted in hidden-sector SUSY
breaking models.
In Fig. \ref{fig:OmXmp100GeV} we show 
the contour plot of $\Omega_X$ with various constraints
for $m_X = 10^{13}$ GeV and $m_\phi = 100$ GeV.
Such heavy moduli are severely constrained not to destroy 
or overproduce light elements synthesized by the big bang
nucleosynthesis. It has been found that 
$(\Omega_\phi)_{BB} h^2 $ and $(\Omega_\phi)_{TI} h^2$ should 
be less than about $10^{-5}$ \cite{Holtman-Kawasaki-Kohri-Moroi}.
Fig. \ref{fig:OmXmp100GeV} shows that in order to solve the moduli problem,
the thermal inflation requires $M_\ast$ higher than in the previous
case,  which results in more dilution of $X$,
and hence we obtain $\Omega_X h^2 \sim 10^{-9}$--$10^{-7}$ for 
$m_X = 10^{13}$ GeV.

Although we have only considered the case $n=1$,
similar discussions also hold in the higher $n$ case,
except that the scale of $M_\ast$ becomes higher.

\begin{figure}[t]
    \centerline{\psfig{figure=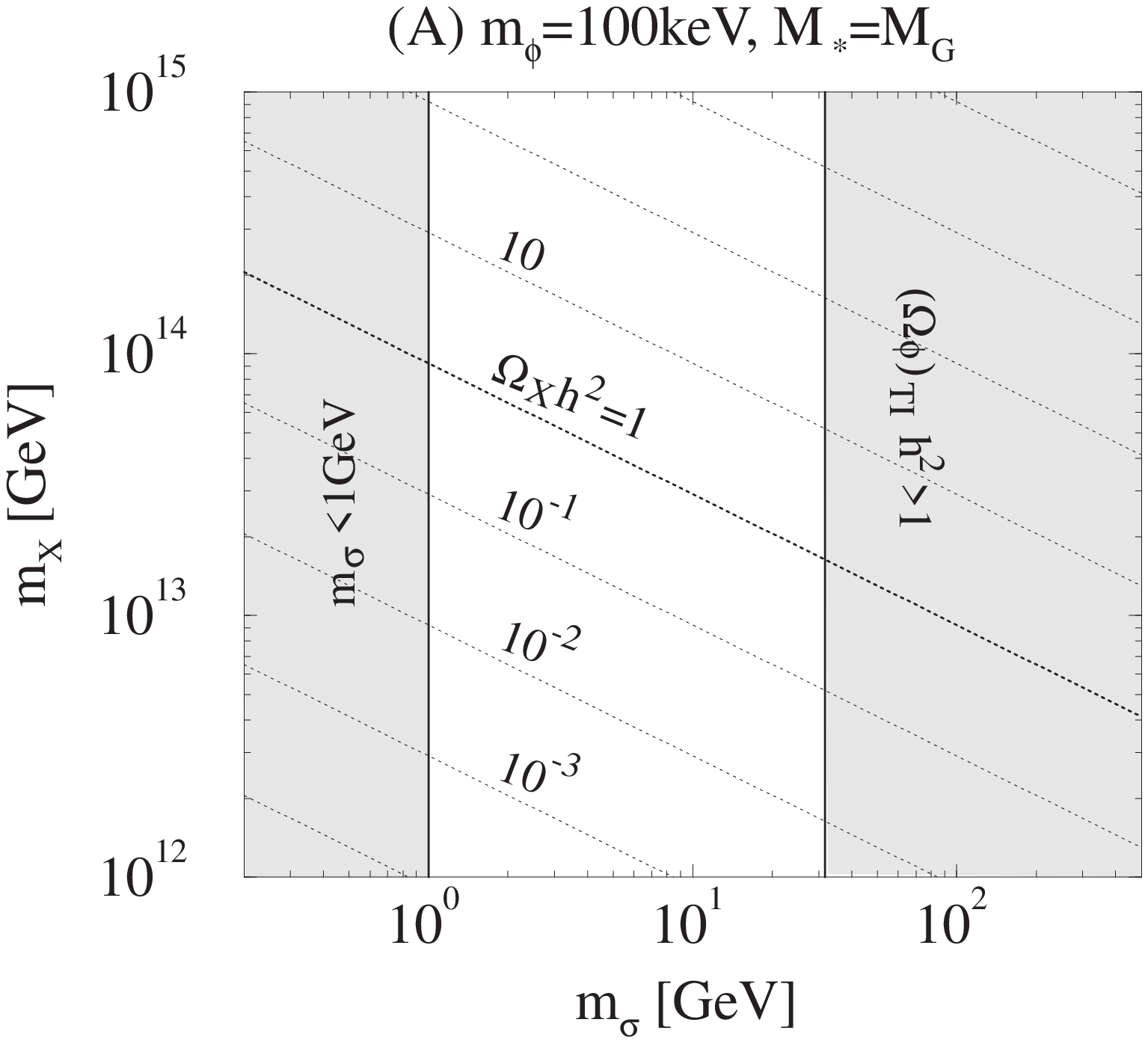,width=10cm}}
    \centerline{\psfig{figure=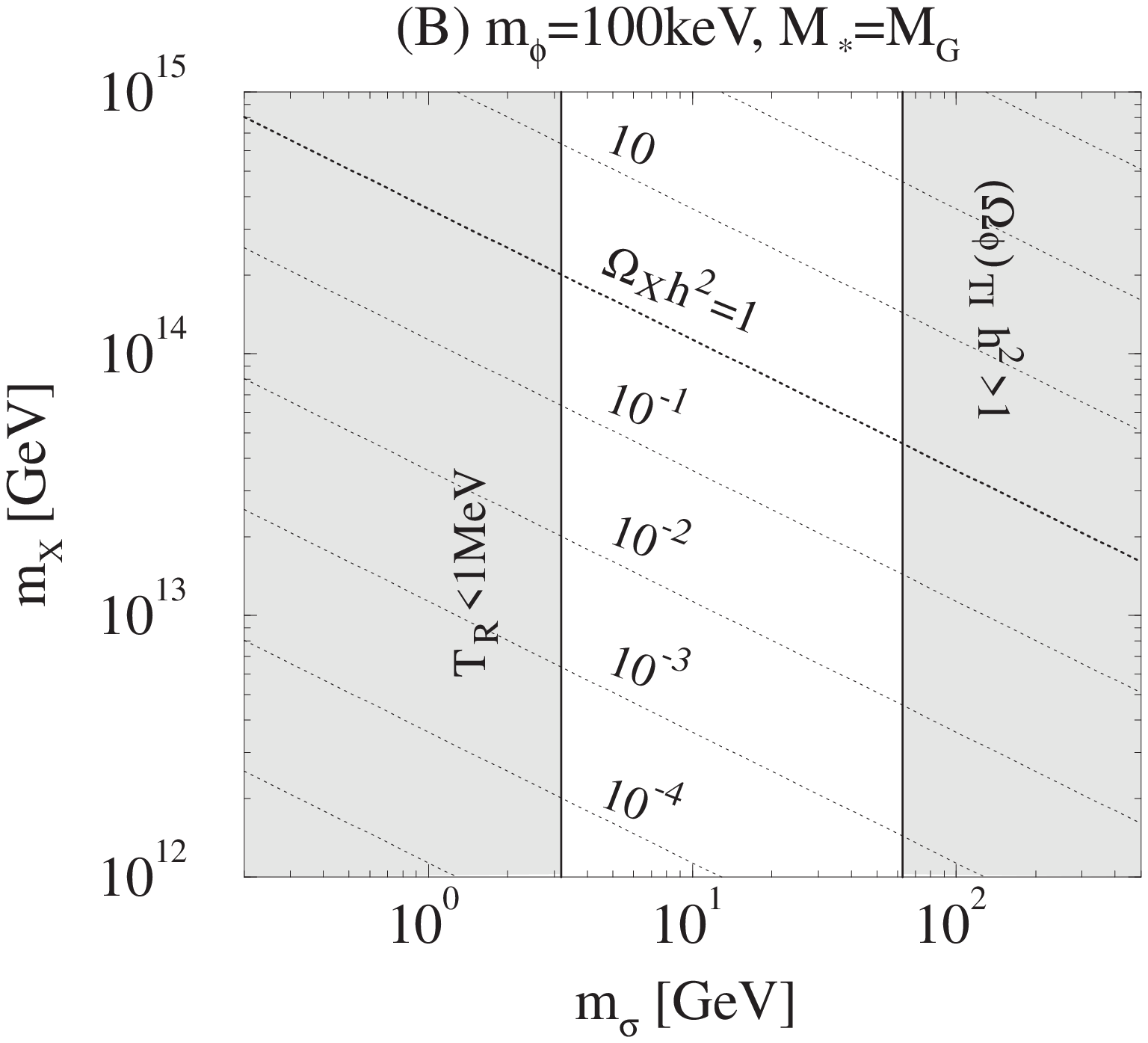,width=10cm}}
\caption{
The contour lines of the relic abundance ($\Omega_X h^2$)
of the superheavy particle $X$ 
in the presence of the thermal inflation 
with $n=1$ for the cases that a flaton $\sigma$ 
decays mainly into two gluons (A)
and that a flaton $\sigma$ decays mainly into two photons (B).
We take the cutoff scale as 
$M_\ast = M_G \simeq 2.4 \times 10^{18}$ GeV and
$m_\phi = 100$ keV.
The contour lines are represented by the dotted lines.
(The contour line of $\Omega_X h^2$=1 is represented by the thick
dotted line.)
We show the lower bound on the flaton mass 
$m_\sigma > 1$ GeV to open the gluon decay channel (A) 
and $m_\sigma > 3.4$ GeV from $T_R >$ 1 MeV (B),
and also the upper bound on $m_\sigma$ 
from $(\Omega_\phi)_{TI} h^2 <1$
by the thick solid lines.
}
    \label{fig:OmXmp100keVMst1D19}
\end{figure}

In the present analysis 
we have taken the cutoff scale of the thermal inflation model
$M_\ast$ as a free parameter.
However,
it is natural to choose it as the gravitational scale, i.e.,
$M_\ast \sim M_G$.
If it is the case,
the thermal inflation dilutes the string moduli
sufficiently only if their masses are $m_\phi \simeq 10^{-1}$keV
--$1$ MeV.
Moreover, as shown in Fig. \ref{fig:OmXmp100keVMst1D19},
the dark matter density of the superheavy particle of mass 
$m_X \simeq 10^{12}$--$10^{14}$ GeV
becomes $\Omega_X h^2 \sim10^{-4}$--1, 
which is just the mass region required to explain 
the observed ultra high-energy cosmic rays.
The required long lifetime of the $X$ particle
may be explained by discrete gauge symmetries
\cite{Hamaguchi-Nomura-Yanagida}
or by compositeness of the $X$ particle
\cite{Long-Lived}.

In this paper we have shown that 
the thermal inflation provides indeed
a new possibility that 
the superheavy $X$ particle may form a part of 
the dark matter in the present universe.
However, it gives rise to a
new problem, since it also dilutes the primordial baryon asymmetry
significantly.
Because the reheating temperature of the thermal inflation 
should be quite low $T_R \sim 1$--10 MeV 
to dilute $\phi$ and $X$ sufficiently,
the electroweak baryogenesis does not work.
However, as shown in Ref. \cite{Gouvea-Moroi-Murayama},
the Affleck-Dine mechanism \cite{Affleck-Dine} may produce 
enough baryon asymmetry even with tremendous entropy production 
due to the thermal inflation,
if the moduli are light ($m_\phi \lesssim 1$ MeV)
\cite{Kawasaki-Yanagida-xray}.
Therefore, in the present scenario, 
the light moduli predicted in gauge-mediated SUSY breaking
models are favored.
\acknowledgments
This work was partially supported by the Japan Society for
the Promotion of Science (TA) and
``Priority Area: Supersymmetry and Unified Theory
of Elementary Particles($\sharp$707)''(MK,TY).


\end{document}